\documentclass[twocolumn,showkeys,showpacs,prl,superscriptaddress]{revtex4}
\usepackage{amssymb}
\usepackage{amsmath}
\usepackage{graphicx}
\usepackage{color}
\usepackage{hyperref}
\usepackage[normalsize]{subfigure}
\usepackage{amssymb}
\usepackage{enumerate}
\usepackage{amsthm,amsfonts,amsmath}
\usepackage{bm}
\usepackage{graphicx}
\usepackage{graphics,epsf,bbold}
\usepackage{siunitx}
\usepackage{color}
\usepackage{xspace}
\usepackage{accents}

\usepackage[T1]{fontenc}
\usepackage[english]{babel}
\usepackage{siunitx}
\usepackage{tensor}
\usepackage{xspace}
\usepackage{bm,bbm}
\usepackage{color}
\usepackage{accents}
\usepackage{setspace}
\usepackage{mathtools}
\usepackage[normalem]{ulem}
\usepackage[capitalize]{cleveref}
\usepackage{soul}
\hypersetup{bookmarks={false}}

\begin{document}

\title{Fluidization and wall slip of soft-glassy materials by controlled surface roughness}

\author{Ladislav Derzsi}
\email{ladislav.derzsi@unipd.it}
\affiliation{Dipartimento di Fisica e Astronomia ``G. Galilei'' - DFA and Sezione CNISM,Universit\`a di Padova,Via Marzolo 8, 35131 Padova, Italy\\}
\author{Daniele Filippi}
\affiliation{Dipartimento di Fisica e Astronomia ``G. Galilei'' - DFA and Sezione CNISM,Universit\`a di Padova,Via Marzolo 8, 35131 Padova, Italy\\}
\author{Giampaolo Mistura}
\affiliation{Dipartimento di Fisica e Astronomia ``G. Galilei'' - DFA and Sezione CNISM,Universit\`a di Padova,Via Marzolo 8, 35131 Padova, Italy\\}
\author{Matteo Pierno}
\email{matteo.pierno@unipd.it}
\affiliation{Dipartimento di Fisica e Astronomia ``G. Galilei'' - DFA and Sezione CNISM,Universit\`a di Padova,Via Marzolo 8, 35131 Padova, Italy\\}
\author{Matteo Lulli}
\affiliation{Dipartimento di Fisica, Universit\`a di Roma  ``Tor Vergata'' and INFN, Via della Ricerca Scientifica, 1 - 00133  Roma, Italy\\}
\author{Mauro Sbragaglia}
\email{sbragaglia@roma2.infn.it}
\affiliation{Dipartimento di Fisica, Universit\`a di Roma  ``Tor Vergata'' and INFN, Via della Ricerca Scientifica, 1 - 00133  Roma, Italy\\}
\author{Massimo Bernaschi}
\affiliation{Istituto per le Applicazioni del Calcolo CNR, Via dei Taurini, 9 - 00185  Roma, Italy\\}
\author{Piotr Garstecki}
\affiliation{Institute of Physical Chemistry, Polish Academy of Sciences, Kasprzaka 44/52, 01-224  Warsaw, Poland\\}

\date{\today}

\begin{abstract}
The motion of soft-glassy materials (SGM) in a confined geometry is strongly impacted by surface roughness. However, the effect of the spatial distribution of the roughness remains poorly understood from a more quantitative viewpoint. Here we present a comprehensive study of concentrated emulsions flowing in microfluidic channels, one wall of which is patterned with micron-size equally spaced grooves oriented perpendicularly to the flow direction. We show that roughness-induced fluidization can be quantitatively tailored by systematically changing both the width and separation of the grooves. We find that a simple scaling law describes such fluidization as a function of the density of grooves, suggesting common scenarios for droplet trapping and release. Numerical simulations confirm these views and are used to elucidate the relation between fluidization and the rate of plastic rearrangements. 
\end{abstract}


\maketitle

Controlling the slip and flow of soft-glassy materials (SGM) at the microscale is crucial for food and pharmaceutical processing, and for micro-manufacturing~\cite{kunii2013fluidization,meeker2004slip,denkov2008viscous,tabilo2005rheology}. SGM include concentrated emulsions, gels, foams, pastes, and exhibit a complex, non-linear rheology~\cite{Larson,Coussot,FOAM}: they behave like elastic solids unless a stress large enough, known as the yield stress $\sigma_{\tiny\mbox{Y}}$, is applied. Above $\sigma_{\tiny\mbox{Y}}$ SGM flow like non-Newtonian liquids. This solid-to-liquid transition and the corresponding flowing properties have been widely studied~\cite{Mason99}, but still pose a series of challenging questions, relevant both for applications~\cite{Angell,mason1996yielding,scheffold2014jamming} and for a better understanding of the statistical mechanics of SGM~\cite{FalkLanger98,Varnik1,Varnik2,Pouliquen09,KEP09,Sollich3,Jop12,Weeks15,ourSM16}. Recent studies~\cite{Goyon08,Goyon10,Seth12,Mansard13,Mansard14,Bonn15,Katgert10,ourJFM15,Geraud13} showed that their flow bahavior is characterized by ``non-locality''~\cite{Goyon08,Goyon10}, meaning that the relation between the local stress $\sigma$ and the local shear rate $\dot{\gamma}$ cannot be explained with a unique master curve. This non-local behaviour depends on both confinement and surface roughness~\cite{Goyon10,Mansard14,Bonn15}, and it is ascribed to the presence of plastic rearrangements~\cite{Goyon08,Goyon10}, i.e. topological changes in the micro-structural configurations. These take place whenever the material cannot sustain the accumulated stress, then it undergoes an irreversible deformation and releases the excess stress in the form of elastic waves. The range of such perturbation introduces a new length, named ``cooperativity length'' $\xi$~\cite{Goyon08}, which is typically on the order of a few diameters of micro-structural constituents (i.e. droplets for emulsions~\cite{Goyon08,Goyon10}, bubbles for foams~\cite{Katgert10,ourJFM15}, blobs for gels~\cite{Geraud13}, etc). 
Although the cooperativity length becomes relevant at the jamming point of SGM~\cite{Nagel98}, it has been sharply argued that $\xi$ is fundamentally different from the characteristic legnth that describes dynamical heterogeneities involved in spontaneous fluctuations~\cite{berthier2005direct,CommReply_Prl.104.169602,ikeda2012unified}.\\
Recently, many theoretical studies have been put forward in the recent years to account for these non-local effects~\cite{FalkLanger98,Pouliquen09,KEP09,Sollich3,ourSM16}. One of them, the kinetic elasto-plastic (KEP) model~\cite{KEP09}, explores the effects of cooperativity on the fluidity field $f=\dot{\gamma}/\sigma$, i.e. the inverse viscosity for the system. An important result of KEP relates the fluidity field with the rate $\Gamma$ of plastic rearrangements
\begin{equation}\label{eq:rate}
\Gamma= \frac{d N_{\tiny\mbox{plastic}}}{dt}
\end{equation}
%
The relation between the rate of plastic rearrangements and the flow profiles is highly non-trivial~\cite{Bouzid15}, and with the exception of a few studies~\cite{Mansard13,ourEPL16,ourJFM15}, has not been researched in detail. \\
Experimental studies report that surface roughness can change the material structural properties, trigger and promote rearrangements, and describe this scenario by ad-hoc parameters, e.g. the wall fluidization~\cite{Goyon08,Goyon10,Bonn15}. Such wall fluidization was generically found to be larger for rough walls than for smooth walls~\cite{Goyon08,Goyon10}, but only few studies provided quantitative insights into how a systematic change of the roughness affects it. Mansard {\it et al.}~\cite{Mansard14} showed that both slippage and wall fluidization depend non-monotonously on the height of the roughness. However, the effect of the roughness geometry on the enhancement of the plastic activity has still to be properly addressed.\\
In this letter we present a comprehensive experimental and numerical study of concentrated emulsions flowing in microchannels, and provide the first direct evidence that the roughness-induced fluidization follows precise scaling laws in terms of the density of rough elements. Thus, it can be quantitatively controlled and predicted. This is understood in terms of plastic activity: as droplets encounter the rough elements, either by flowing into the gaps or by hitting an obstacle, they suddenly change their local velocity and induce an irreversible rearrangement of their neighbors. The link between fluidization and plastic rearrangements is then directly tested~\cite{Bouzid15}. \\
%
%
\begin{figure}[t!]
\includegraphics[scale=1.0]{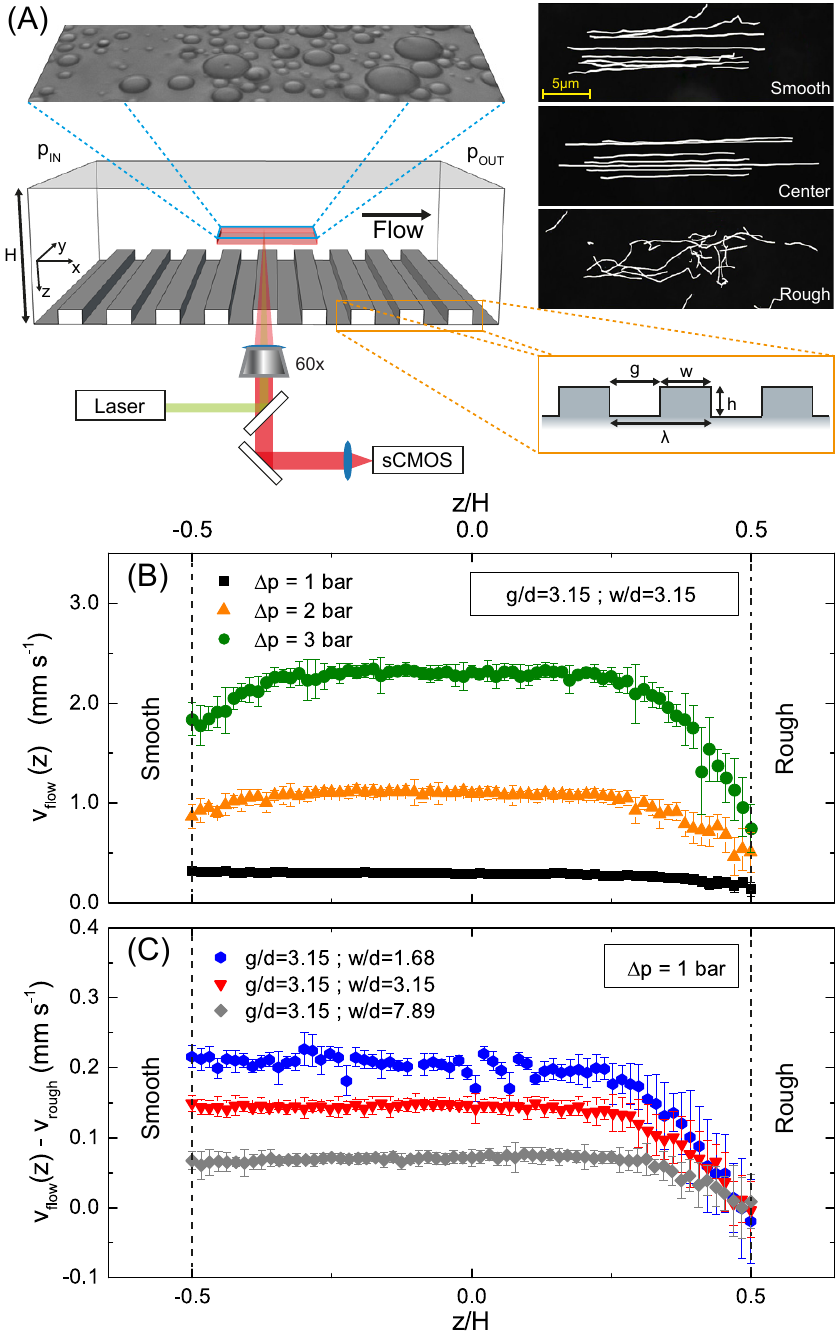}
\caption{(A): sketch of the experimental setup used for flow visualization of dense emulsions. The flow is driven by a pressure drop $\Delta p=p_{\tiny\mbox{IN}}-p_{\tiny\mbox{OUT}}$ through a microfluidic channel of height $H$, one wall of which is patterned with a controlled roughness. 
%
%
Black backgrounded snapshots show the trajectories of fluorescent nanoparticles suspended in the continuous phase of the emulsion, flowing at different channel height $z$: close to the rough wall (bottom snap), in the center of the channel (middle snap, corresponding to the plug region) and close to the smooth wall (top snap). Flow profiles $\mbox{v}_{\tiny\mbox{flow}}(z)$, the origin of $z$ being at the center of the channel height, are reported either at fixed roughness for different pressure drops (B) or at fixed pressure drop for different surface roughness (C). In the latter case it has been subtracted the slip velocity on the rough wall $\mbox{v}_{\tiny\mbox{rough}}$.
}
\label{fig:1}
\end{figure}
%
We studied the flow behavior of a concentrated emulsion in a home-made microfluidic device prepared by multilayer photolithography using SU8 photoresist on glass~\cite{ferraro2012morphological,brigo2008water}. The microfluidic channels had a width $W$ of 4 mm, height $H$ of 220 $\mu$m and length $L$ of 4.5 cm. Controlled roughness was provided by an array of rectangular posts of width $w$, gap $g$ and period $\lambda=w+g$, extending over the whole width of the channel (see bottom zoom of~\cref{fig:1}-A). The posts were orientated perpendicularly to the direction of the flow and had heights of $h\approx 2.3\;\mu$m. Only one face of the channel was patterned ($z=H/2$), the other glass was smooth ($z=-H/2$). In this way it was possible to compare the behavior of the emulsion in proximity of the two different walls. Aqueous solution of polyvinylpyrrolidone was used to make hydrophilic the walls. The concentrated emulsion was prepared of silicone oil (polydimethyl siloxane, $1000~\mbox{mPa} \cdot \mbox{s}$) dispersed in a 50\% aqueous glycerine and stabilized by 1\, wt\% Tetradecyl Trymethyl Ammonium Bromide (TTAB). The concentration of the surfactant was set high enough to prevent coalescence of the droplets, yet low enough to avoid flocculation by depletion~\cite{Goyon08}. The emulsion was optically transparent and non-adhesive. The mean diameter of the emulsion was $d=4.75\;\mu$m with a polydispersity index (coefficient of variance) of 0.61, and the volume fraction of the disperse phase was $\Phi=0.875$. We measured the (bulk) rheological properties of the emulsion using a rotational rheometer with cone-plate geometry. The flow curve is well described by the Herschel-Bulkley model (see Sec.~I of the electronic supplementary information, ESI). Finally, we seeded the continuous phase of the emulsion with a diluted suspension (0.002\,wt\%) of fluorescent nanoparticles (size of $\approx0.2\;\mu$m) to measure the flow profiles using Particle Tracking Velocimetry (PTV) methods~\cite{Adrian91}. 
%
%
Tracers were illuminated with a $532$ nm laser beam and imaged via an inverted, motorized, microscope equipped with a sCMOS camera (see Sec.~I of the ESI).
We recorded $z$-stacks of images
%
%
with steps of $\approx0.5$ droplet diameter $d$. 
%
%
%
{Trajectories crossing the optical field within the depth of focus of the microscope objective were then acquired. The magnitude of the tracers' velocity $\mbox{v}_{\tiny\mbox{flow}} (z)$ at the stack $z$ was measured by averaging over hundreds of different tracks, collected by sampling different region of interest (ROIs) throughout the channel. 
In each flow profile a difference of the wall slip velocities on the rough ($z= H/2)$ and the smooth ($z=- H/2)$ wall 
\begin{equation}\label{eq:slip}
\Delta \mbox{v}_{\tiny\mbox{slip}}=\mbox{v}_{\tiny\mbox{smooth}}-\mbox{v}_{\tiny\mbox{rough}}
\end{equation} 
emerges clearly, being $\mbox{v}_{\tiny\mbox{smooth}} > \mbox{v}_{\tiny\mbox{rough}}$ (see~\cref{fig:1}-B,C). By increasing the applied pressure drop, the flow profiles partially recover the classical Poiseuille profile, yet they still display a plug region in the center, characteristic of yield stress fluids (low shear regions where $\sigma(z)<\sigma_{\tiny\mbox{Y}}$). Additionally, by increasing the pressure drop, the difference of slip velocities
increases (\cref{fig:1}-B). 
Figure~\ref{fig:1}-C shows the flow profiles shifted by $\mbox{v}_{\tiny\mbox{rough}}$ for a fixed pressure drop.
%
%
By keeping the width of the gap $g$ fixed and changing the widths of the post $w$, we observed a systematic evolution in the flow profiles: increasing $\lambda$ lead to a smaller fluidization (i.e. velocity gradient) and consequently smaller $\Delta \mbox{v}_{\tiny\mbox{slip}}$. We note that the same effect could be observed if $w$ was kept fixed and $g$ changed, since both for $g/d \gg 1$ and $w/d \gg 1$ we must recover a vanishing $\Delta \mbox{v}_{\tiny\mbox{slip}}$. From the micro-mechanical point of view, the observed decrease of $\mbox{v}_{\tiny\mbox{rough}}$ (in respect to $\mbox{v}_{\tiny\mbox{smooth}}$) can be accounted for the increased number of plastic rearrangements~\cite{Goyon08,Goyon10}: droplets encountering the rough elements (gaps, posts) change their speed and often also their direction causing a scramble with their neighbors, whereas their falter on the smooth wall is rather occasional and almost absent in the plug region (see \cref{fig:1}-A, movies M1 (experiments) and M2 (simulations) of the ESI). \\
\begin{figure}[th!]
\centering
\includegraphics[scale=1]{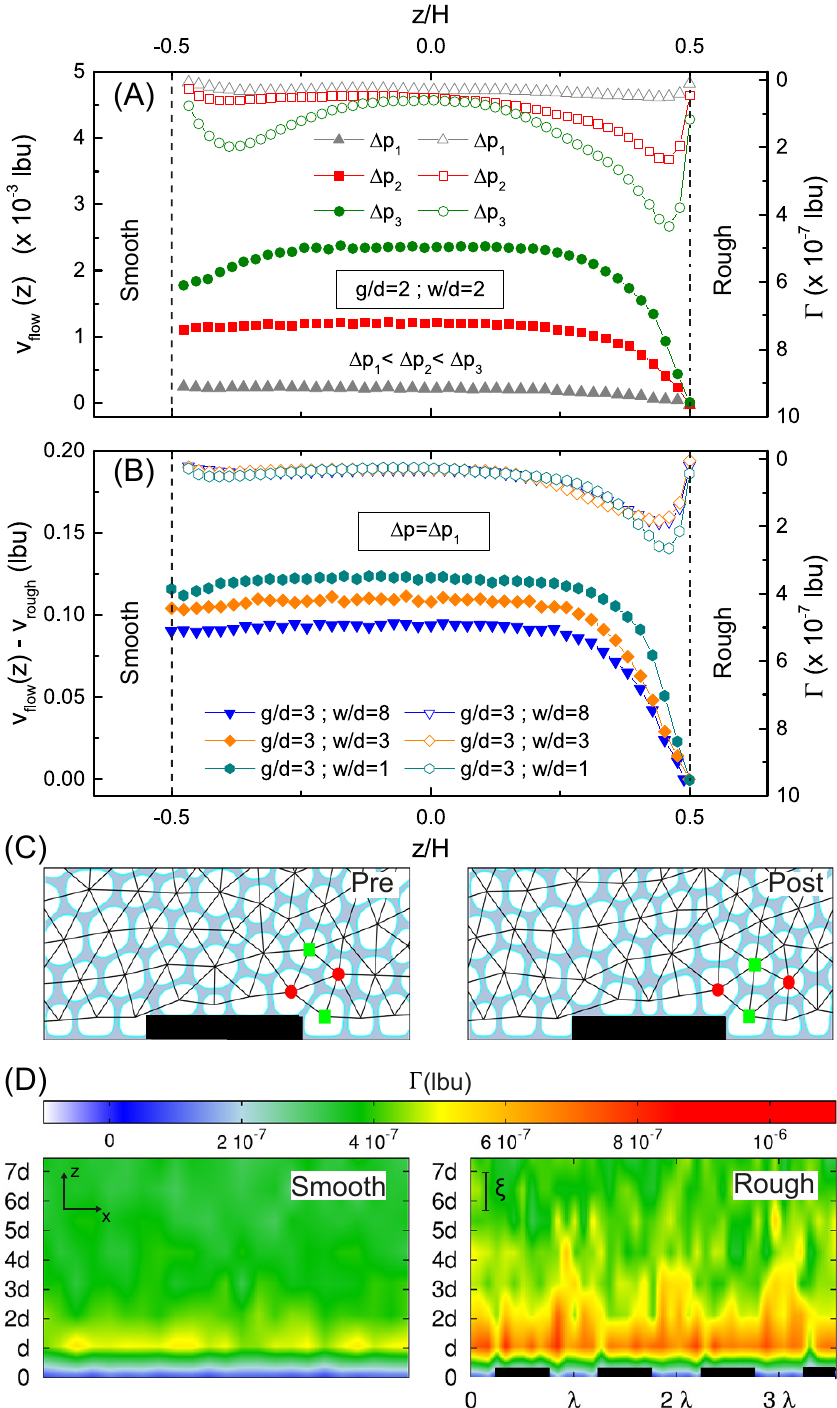}\\
\caption{Insights on flow profiles and rate of plastic rearrangements $\Gamma$ \eqref{eq:rate} from numerical simulations based on the lattice Boltzmann methods~\cite{ourSM12,ourJFM15,ourEPL16}. (A) velocity profiles and the associated $\Gamma$ for a fixed roughness and increasing pressure drops $\Delta p$. Data at fixed  $\Delta p$ for different roughness are also displayed (B). Plastic rearrangements are identified by comparing two consecutive configurations of Delaunay triangulations built from the centers of mass of the droplets (C). The spatial distribution of $\Gamma$ in the sheared layer close to a smooth and rough wall is also displayed (D). The cooperativity length $\xi$ is indicated with a vertical bar. Results are reported in lattice Boltzmann units (lbu).}
\label{fig:2}
\end{figure}
%
Numerical simulations of a model emulsion based on the lattice Boltzmann methods corroborated this physical picture. The numerical model is in continuity with a number of works by some of the authors~\cite{ourSM12,ourJFM15,ourEPL16} (see Sec.~II of ESI). The simulations allowed to access simultaneously the flow profiles and plastic rearrangements via the analysis of the Delaunay triangulation~\cite{Delaunay}. To this aim we applied a novel procedure to detect rearrangements in geometries with arbitrary complex boundaries~\cite{ourGPU16}. This made it possible to study the phenomena related to droplets plasticity with unprecedented statistics~\cite{ourJFM15,ourEPL16}. Shortly, we compared two consecutive configurations by using the corresponding Delaunay triangulations built starting from the centers of mass of the droplets. The triangulation provides the nearest neighbors of each droplet. A topological change takes place every time a link between two droplets disappears, for boundary events, and a new one appears for bulk events (see~\cref{fig:2}-C).  The observed experimental behaviours illustrated in \cref{fig:1} were actually well reproduced by our numerical simulations, as reported in~\cref{fig:2}-A and~\cref{fig:2}-B, where the corresponding rates of plastic rearrangements are also displayed. These figures clearly show that both the asymmetry of the profiles at fixed roughness (\cref{fig:2}-A) and the different fluidization properties at fixed pressure drop (\cref{fig:2}-B) have a direct correspondence to the rate of plastic activity \eqref{eq:rate}, with the latter being more pronounced at high velocity gradient. We also remark~\cite{ourSM12,ourEPL16} that the cooperativity scenario~\cite{Goyon08,Goyon10} underlies these observations. To highlight this, we performed numerical simulations of a Couette cell at fixed shear stress~\cite{ourEPL16} (we did not test it experimentally, because there exist specific studies~\cite{Seth12,Bonn15}). Indeed, in a Couette cell, we had the possibility to measure directly the effects of cooperativity in the deviations from linearity of the velocity profiles~\cite{Goyon10}, as described in Sec.~II of the ESI. These simulations reveal that roughness induces a substantial increase of the wall fluidity (see Fig.~S2 of the ESI) which decays towards the bulk fluidity according to the theoretical predictions~\cite{Goyon10}. The measured value of the cooperativity length was $\xi \approx 1.6\,d$, in agreement with other existing observations~\cite{Goyon08,Goyon10,Bonn15}. It is noteworthy to remark that our findings support the idea that the cooperativity length regulates the protrusion into the channel of the plastic activity triggered by the rough wall~\cite{Goyon08,Goyon10,Bonn15}, as evident from~\cref{fig:2}-D that reports the spatial distribution of plastic rearrangements in the sheared layer close to a smooth and rough walls. For the smooth wall, $\Gamma$ is relatively small and rather homogeneous along the flow ($x$ direction); this contrasts the case of a rough wall, where $\Gamma$ is enhanced, with a periodic modulation dictated by the roughness.\\
One expects that the slip velocity on the smooth wall $\mbox{v}_{\tiny\mbox{smooth}}$ is constant for a given pressure drop and independent of the periodicity of the rough wall on the other side of the channel. On the other hand, surface roughness (depending on its structure) strongly affects plastic rearrangements, thus the slip conditions (see Fig.~S1 in the ESI). Therefore, to further characterize the effect of roughness periodicity and quantify the roughness-induced fluidization we considered $\Delta \mbox{v}_{\tiny\mbox{slip}}$ as a function of $\lambda$. Results show clear evidence that, for a fixed pressure drop, $\Delta \mbox{v}_{\tiny\mbox{slip}}$ scales with the periodicity of the roughness (see \cref{fig:3}-A)
\begin{equation}
\Delta \mbox{v}_{\tiny\mbox{slip}} \sim \lambda^{-1}
\end{equation} 
where the prefactor is found to increase with the applied pressure drop. This may be taken into account by further normalizing $\Delta \mbox{v}_{\tiny\mbox{slip}}$ for a characteristic velocity dependent on the pressure drop. To this aim, we calculated the maximum (plug) velocity $\mbox{v}_{\tiny\mbox{plug}}$ by averaging the measured velocities in the 8-10 droplet diameter region in the center of the channels (plug region). Upon rescaling with $\mbox{v}_{\tiny\mbox{plug}}$ (see \cref{fig:3}-B), the individual scaling curves nicely collapse into a single master curve (see \cref{fig:3}-B). We note that, since plastic rearrangements and cooperativity effects depend on the volume fraction~\cite{Goyon08,Goyon10}, this effect may be less pronounced at lower concentrations of the emulsion.\\
Finally, \cref{fig:4} shows the relation between the numerical $\Delta \mbox{v}_{\tiny\mbox{slip}}$ and the rate of plastic rearrangements. Again simulations in a Couette cell for both smooth and rough channels were performed. The imposed (nominal) shear was kept constant by applying a fixed velocity at the wall in $z=-H/2$, while changing the nature (i.e. smooth or rough) of the wall in $z=+H/2$ (see Sec.~II of the ESI). The results of numerical simulations confirm the observed experimental scaling $\Delta \mbox{v}_{\tiny\mbox{slip}} \sim \lambda^{-1}$ (inset of \cref{fig:4}). More important, we computed the increase of the rate of plastic rearrangements (due to the rough wall) in respect to the case with a smooth wall, $\Gamma_{\tiny\mbox{TOT}}=\frac{1}{4}\int (\Gamma_{\tiny\mbox{rough}}(\vec{x})-\Gamma_{\tiny\mbox{smooth}}(\vec{x}))\, d\vec{x}$, and observed a linear relation between $\Gamma_{\tiny\mbox{TOT}}$ and $\Delta \mbox{v}_{\tiny\mbox{slip}}$. This result is highly non trivial~\cite{Bouzid15}: it is not only a further confirmation that the roughness-induced fluidization is due to plastic rearrangements, but it also supports one of the key results of KEP~\cite{KEP09} which we carefully verified and tested in presence of rough walls with variable widths and gaps.\\
\begin{figure}[t!]
\centering
\includegraphics[scale=1.0]{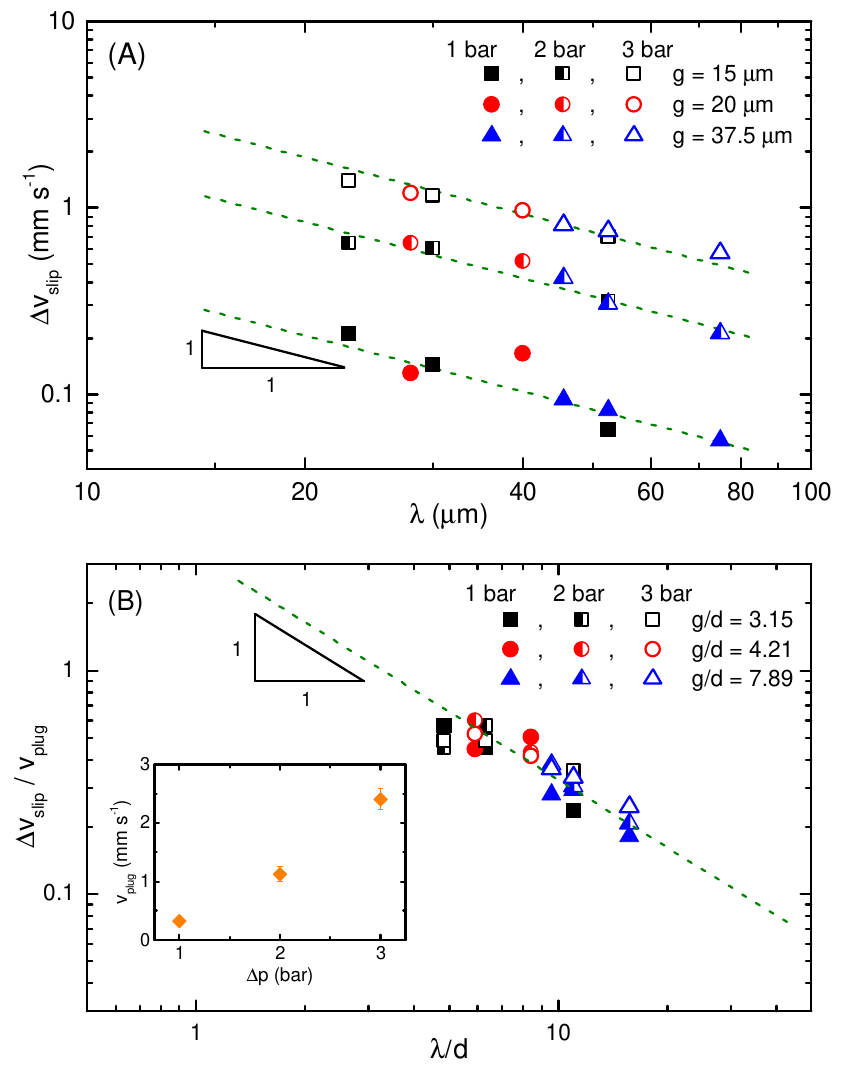}\\
\caption{(A): Difference of the slip velocities~\eqref{eq:slip} as a function of the roughness periodicity $\lambda = w+g$ for different pressure drops $\Delta p$ and different gaps $g$. Dashed lines are guides for the eyes with slope $-1$. (B): Difference of the slip velocities, normalized by the maximum (plug) velocity $\mbox{v}_{\tiny\mbox{plug}}$, as a function of the wall roughness periodicity $\lambda$ in units of the mean droplet diameter $d$. Dashed line is a guide for the eyes with slope $-1$. Inset shows the dependence of $\mbox{v}_{\tiny\mbox{plug}}$ with $\Delta p$.}
\label{fig:3}
\end{figure}
%
%
%
\begin{figure}[h!]
\centering
\includegraphics[scale=1.0]{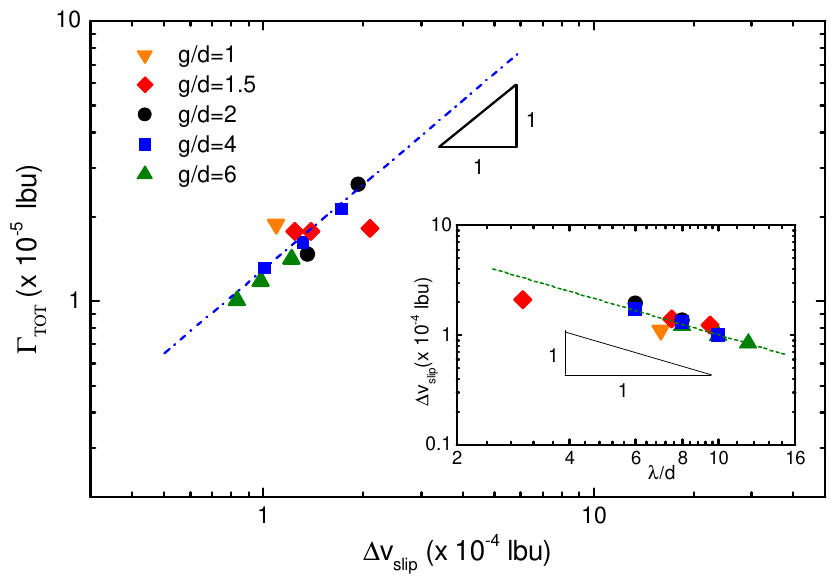}\\
\caption{Lattice Boltzmann simulation results for the increase of the rate of plastic rearrangements induced by the rough wall, $\Gamma_{\tiny\mbox{TOT}}$ (see text for details). $\Gamma_{\tiny\mbox{TOT}}$ is plotted as a function of normalized slip variation $\Delta \mbox{v}_{\tiny\mbox{slip}}$ \eqref{eq:slip}. Dashed-dotted line is a guide for the eyes with slope $1$. The inset reports the slip variation as a function of the dimensionless period $\lambda/d$. Dashed line is a guide for the eyes with slope $-1$. Data are reported in lattice Boltzmann units (lbu).}\label{fig:4}
\end{figure}
In summary, we have observed both experimentally and numerically that roughness-induced fluidization, i.e. the enhanced plastic activity induced by rough walls~\cite{Goyon08,Goyon10}, can be tailored upon designing a micro-structured surface to control the motion of SGM in microfluidic channels. This is due to the modulated plastic activity triggered by the roughness, and results in a simple, yet non trivial scaling relation for the fluidization as a function of the roughness periodicity. This scenario is observed for periodicity patterns spanning more than one decade in droplets diameters and for gaps larger than the cooperativity length ($\sim 2-3\,d$).\\
\\
The research leading to these results has received funding from the European Research Council under the European Community's Seventh Framework Programme (FP7/2007-2013)/ERC Grant Agreement No. 279004. The authors also thank Giovanni Lucchetta and Julie Goyon for useful advices. PG acknowledges founding by the European Research Council Starting Grant 279647.


\bibliography{refs}

\end{document}